\documentclass[sn-nature]{sn-jnl}

\usepackage[a4paper, margin=1in]{geometry}
\usepackage{graphicx}%
\usepackage{multirow}%
\usepackage{amsmath,amssymb,amsfonts}%
\usepackage{amsthm}%
\usepackage{mathrsfs}%
\usepackage[title]{appendix}%
\usepackage{xcolor}%
\usepackage{textcomp}%
\usepackage{manyfoot}%
\usepackage{booktabs}%
\usepackage{algorithm}%
\usepackage{algorithmicx}%
\usepackage{algpseudocode}%
\usepackage{listings}%
\usepackage[version=3]{mhchem} 
\usepackage[english]{babel}
\usepackage{units}
\usepackage{gensymb}
\usepackage{setspace}
\usepackage{lineno}

\begin{document}

\title[Room temperature nonlocal detection of charge-spin interconversion in a topological insulator]{\textbf{Room temperature nonlocal detection of charge-spin interconversion in a topological insulator}}


\author[1]{\fnm{Anamul Md.} \sur{Hoque}}
\equalcont{These authors contributed equally to this work.}

\author[1]{\fnm{Lars} \sur{Sjöström}}
\equalcont{These authors contributed equally to this work.}

\author[1]{\fnm{Dmitrii} \sur{Khokhriakov}}

\author[1]{\fnm{Bing} \sur{Zhao}}

\author*[1]{\fnm{Saroj P.} \sur{Dash}}\email{saroj.dash@chalmers.se}

\affil[1]{\orgdiv{Department of Microtechnology and Nanoscience}, \orgname{Chalmers University of Technology}, \orgaddress{\city{Gothenburg}, \postcode{SE-41296}, \country{Sweden}}}


\abstract{
Topological insulators (TIs) are emerging materials for next-generation low-power nanoelectronic and spintronic device applications. TIs possess non-trivial spin-momen\-tum locking features in the topological surface states in addition to the spin-Hall effect (SHE), and Rashba states due to high spin-orbit coupling (SOC) properties. These phenomena are vital for observing the charge-spin conversion (CSC) processes for spin-based memory, logic and quantum technologies. Although CSC has been observed in TIs by potentiometric measurements, reliable nonlocal detection has so far been limited to cryogenic temperatures up to \emph{T} = 15 K. Here, we report nonlocal detection of CSC and its inverse effect in the TI compound \ce{Bi_{1.5}Sb_{0.5}Te_{1.7}Se_{1.3}} at room temperature using a van der Waals heterostructure with a graphene spin-valve device. The lateral nonlocal device design with graphene allows observation of both spin-switch and Hanle spin precession signals for generation, injection and detection of spin currents by the TI. Detailed bias- and gate-dependent measurements in different geometries prove the robustness of the CSC effects in the TI. These findings demonstrate the possibility of using topological materials to make all-electrical room-temperature spintronic devices.
}




\maketitle

\section*{Introduction}

Topological insulators (TIs) have attracted significant attention in condensed matter physics and information technology because of their electronic band structure with topologically protected electronic states. \cite{Tokura2017EmergentMaterials,Qi2011TopologicalSuperconductors,Sierra2021VanOpto-spintronics} Their insulating bulk bands have Rashba spin-split bands, and their topological surface states (TSS) create gapless metallic Dirac states with helical spin textures induced by strong spin-orbit coupling (SOC). \cite{Hasan2010Colloquium:Insulators,Xia2009ObservationSurface,Hsieh2009ARegime,Lei2022GrapheneDevices,Chakraborty2022ChallengesDevices}
The TSS are characterized by spin-momentum locking (SML), where the electron spin orientation is locked perpendicularly to its momentum, whereas the bulk spin-split bands are spin-polarized as well owing to the Rashba-Edelstein effect (REE).  Such spin-polarized states have a great potential for spintronic technologies, as application of a charge current can create a significant non-equilibrium spin density, providing a large charge-spin conversion (CSC) efficiency. \cite{Wu2021MagneticInsulators,Mellnik2014Spin-transferInsulator,Manchon2019Current-inducedSystems,DC2023ObservationMnPd3} The spin polarization in TIs has previously been utilized to create energy-efficient magnetization dynamics and switching of an adjacent ferromagnet (FM) in a heterostructure via spin–orbit torque (SOT) phenomena. \cite{Mellnik2014Spin-transferInsulator,Fan2014MagnetizationHeterostructure,Khang2018ASwitching,Han2017Room-TemperatureInsulator} The SOT effect in TIs can have contributions from the TSS, quantum confinement of the surface and bulk bands at the Fermi level, and interfacial effects in TI/FM heterostructures. \cite{Valla2012PhotoemissionInsulator,Bonell2020ControlHeterostructures,Li2012MagneticBi2Se3}

The CSC effects in TIs have been investigated using ferromagnetic (FM) tunnel contacts in potentiometric measurements up to room temperature \cite{Li2014ElectricalBi2Se3,Dankert2015RoomInsulators,Ando2014ElectricalBi1.5Sb0.5Te1.7Se1.3,Tang2014ElectricalInsulator,Liu2015Spin-polarizedInsulators,Tian2015ElectricalBi2Te2Se,Li2016DirectStates,Tian2017ObservationBattery} and the competition between the bulk and surface contribution has been evaluated. \cite{Dankert2018OriginInsulators} The spin-switch and Hanle spin precession signals in a nonlocal (NL) measurement geometry unequivocally confirm the manifestation of pure spin current without any charge current contribution. \cite{Han2014GrapheneSpintronics,Avsar2020Colloquium:Materials,Safeer2022ReliabilityValves} In this regard graphene-based hybrid spin-valve devices are useful for NL measurement geometry due to graphene's excellent spin transport properties and ability to combine with other layered materials in van der Waals (vdW) heterostructures. \cite{Khokhriakov2020Two-dimensionalGraphene,Khokhriakov2020RobustBoundaries,Khokhriakov2022MultifunctionalCircuits,Vaklinova2016Current-InducedHeterostructures,Voerman2019Spin-MomentumMeasurements,Zhao2023AHeterostructure} However, reliable NL detection of CSC phenomena in TIs using a graphene spin-valve device has so far been limited to cryogenic temperatures, up to $T = \unit[15]{K}$, limiting its practical applications. \cite{Vaklinova2016Current-InducedHeterostructures}

Here, we demonstrate room temperature CSC and its inverse effect (ICSC) in the layered TI material \ce{Bi_{1.5}Sb_{0.5}Te_{1.7}Se_{1.3}} (BSTS), using a vdW heterostructure with a graphene spin-valve device. The generated spin current in the TI is injected into a chemical vapor deposition (CVD) graphene channel and subsequently detected using a FM contact in a hybrid NL spin-valve device. We take advantage of the lateral NL spin-valve device design with graphene to observe both spin-switch and Hanle spin precession signals, reliably demonstrating the generation, injection and detection of pure spin currents by the TI. Detailed measurements in different device geometries and bias- and gate-dependent studies at room temperature prove the robustness of the room temperature CSC effects in the TI, opening routes for future spintronic device applications.

\section*{Results and Discussion}

\begin{figure}[ht!]
    \centering
    \includegraphics[width=\textwidth]{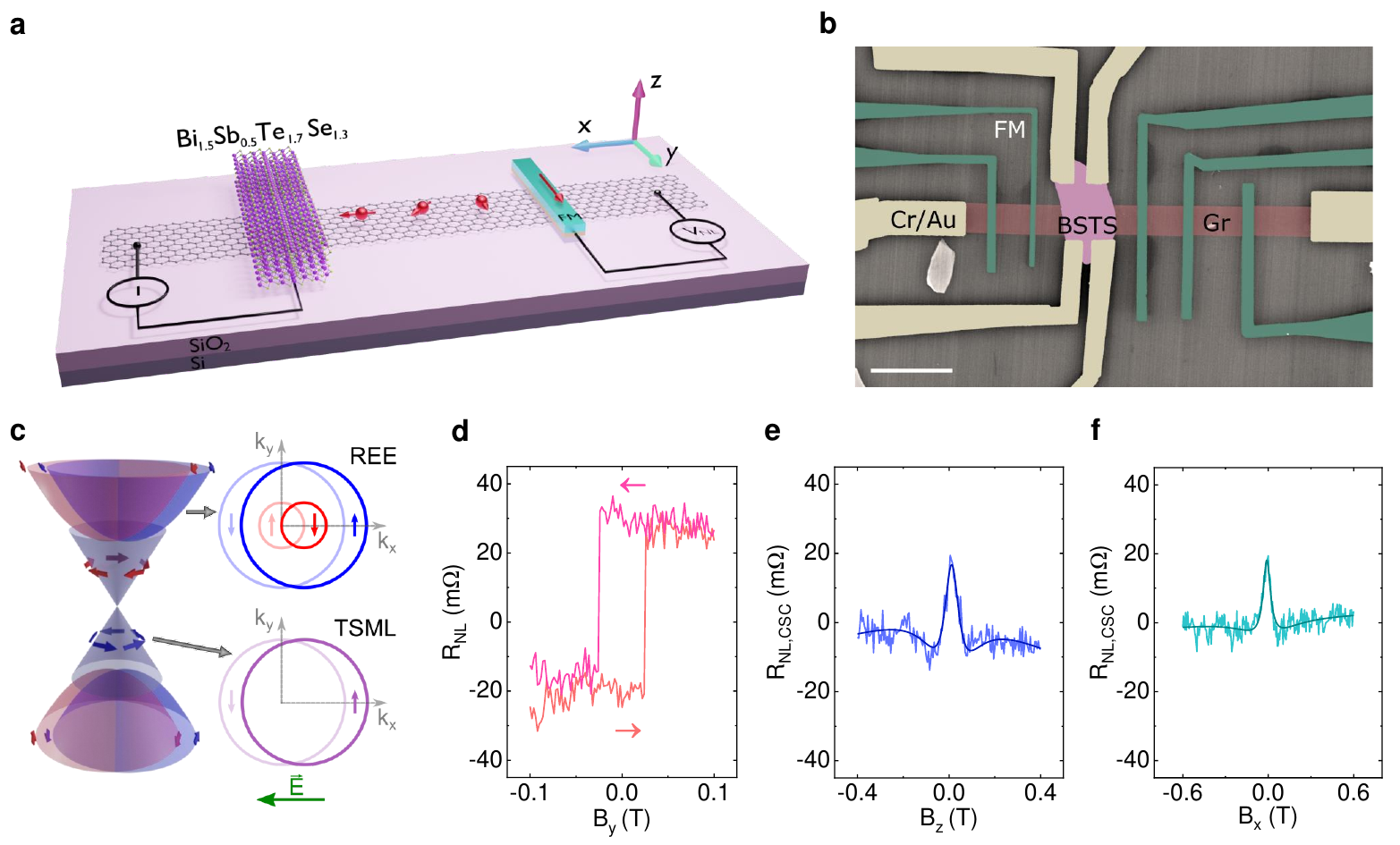}
    \caption{{\footnotesize \textbf{Graphene/BSTS heterostructure device for nonlocal detection of charge-spin conversion at room temperature.} \textbf{(a, b)} Schematic and colored SEM image of a representative graphene/BSTS heterostructure device (Device 2) with NL measurement geometry, with reference nonmagnetic (Cr/Au) and FM (\ce{TiO2}/Co) contacts. The scale bar in the SEM image is $\unit[5]{\mu m}$. \textbf{(c)} Schematic of two CSC mechanisms due to the Rashba-Edelstein effect (REE) and topological spin-momentum locking (TSML) property. \textbf{(d)} Spin-switch signal ($R_{\mathrm{NL}} = V_{\mathrm{NL}}/I$) for spin injection from BSTS with $B_\mathrm{y}$ magnetic field sweep at $V_\mathrm{g} = \unit[30]{V}$ and $I = \unit[-300]{\mu A}$. \textbf{(e)} Hanle spin precession signal ($R_{\mathrm{NL}}$) with $B_\mathrm{z}$ magnetic field sweep, along with the Hanle fitting (solid line) at $V_\mathrm{g} = \unit[30]{V}$ and $I = \unit[-200]{\mu A}$. \textbf{(f)} Hanle spin precession signal ($R_{\mathrm{NL}}$) with $B_\mathrm{x}$ magnetic field sweep at $V_\mathrm{g} = \unit[50]{V}$ and $I = \unit[-300]{\mu A}$, along with Hanle fitting (solid line). The data in (d-f) was measured in Device 1.}}
    \label{fig:1}
\end{figure}

The motivation behind using BSTS is its Fermi level in the band gap, giving rise to a dominant 
surface contribution, \cite{Ren2011OptimizingRegime,Pan2014LowStudy} reportedly $\unit[22]{meV}$ below the conduction band edge as reported by our group. \cite{Dankert2018OriginInsulators} We utilize BSTS and graphene vdW heterostructure-based spin-valve devices to detect CSC effect in the TI. Figures \ref{fig:1}a-b present the device schematic on a Si/\ce{SiO2} substrate and a scanning electron microscope (SEM) image of a fabricated device, respectively. The graphene/BSTS vdW heterostructure devices consist of an exfoliated BSTS flake on CVD graphene. We choose CVD graphene as the spin channel material because it has been shown to exhibit a robust and long-distance spin transport for multifunctional spin-logic operation at room temperature, \cite{Khokhriakov2020Two-dimensionalGraphene,Khokhriakov2020RobustBoundaries,Khokhriakov2022MultifunctionalCircuits} as well as being more suitable than exfoliated graphene for future large-scale device integration. \cite{Han2014GrapheneSpintronics,Khokhriakov2020Two-dimensionalGraphene,Bisswanger2022CVDTemperature} We use nonmagnetic ($\unit[20]{nm}$ Cr/$\unit[90]{nm}$ Au) electrodes as reference contacts and FM ($\unit[\sim 1]{nm}$ \ce{TiO2}/$\unit[90]{nm}$ Co) contacts for spin injection and detection in the graphene channel (see Methods and the Supplementary Information for details). Here, the \ce{TiO2} oxide layer underneath the Co electrodes acts as a tunnel barrier to mitigate  the conductivity mismatch between the graphene channel and FM contacts. \cite{Fert2001ConditionsSemiconductor,Kamalakar2016InversionHeterostructures}

Interestingly, the CSC in TIs at the room temperature can arise due to the bulk spin Hall effect (SHE), the Rashba-Edelstein effect (REE) of the bulk states or of the trivial surface states, and/or the topological spin-momentum locking (TSML) of the TSS. \cite{Sinova2015SpinEffects,Sun2019LargeHeterostructures} A simplified band diagram of BSTS is shown in Figure \ref{fig:1}c, where the Fermi surface has a winding spin texture for both REE and TSML, depending on the Fermi level position. The spin texture winding, in momentum space, will be offset from the equilibrium position due to an applied electric field ($\Vec{E}$), creating a non-equilibrium spin density during CSC (or vice versa during the ICSC effect).

We use a recently developed device geometry and measurement technique to detect CSC in the TI (Figure \ref{fig:1}a), \cite{Zhao2020UnconventionalWTe2,Zhao2020ObservationTemperature,Zhao2020Charge-spinDesign,Hoque2020Charge-spinHeterostructures,Kovacs-Krausz2020ElectricallyBiTeBr,Ontoso2023UnconventionalHeterostructures} where an applied charge current creates a non-equilibrium spin density due to the CSC process in the TI, which is then injected into the graphene spin channel. The diffused spins in the graphene channel is detected as a NL voltage by a remote FM electrode. 
First, spin-switch measurements were performed while sweeping an in-plane magnetic field along the $y$ axis ($B_\mathrm{y}$), which is the easy axis of the FM electrodes. The applied $B_\mathrm{y}$ field switches the magnetization direction of the detector's FM from parallel to antiparallel orientation with respect to the injected spin-polarized electrons from the TI. This results in a change in NL resistance ($\Delta R_{\mathrm{NL}} = \Delta V_{\mathrm{NL}}/I = \unit[42]{m\ohm}$) at room temperature, as shown in Figure \ref{fig:1}d, where $\Delta V_{\mathrm{NL}}$ is the change in measured voltage and $I$ is the applied charge current across the graphene/TI structure in Device 1. Here, the graphene channel length between the TI and the FM contact is $L = \unit[1.84]{\mu m}$.

In order to prove the spin origin of the signal, Hanle spin precession measurements were conducted by sweeping the magnetic field out-of-plane along the $z$ axis ($B_\mathrm{z}$). This causes the in-plane spins to precess and dephase while diffusing along the graphene channel, and reach the FM detector electrode with a finite angle with respect to the contact magnetization direction. \cite{Avsar2020Colloquium:Materials} The Hanle spin precession signals in Figure \ref{fig:1}e-f were obtained by averaging the signals for $+y$ and $-y$ FM detector contact magnetization orientations, $R_{\mathrm{NL,CSC}} = (R_{\mathrm{NL,}\uparrow} - R_{\mathrm{NL,}\downarrow})/2$. The raw data is shown in Supplementary Figure 1b-c. As seen in Figure \ref{fig:1}e, the Hanle signal $R_{\mathrm{NL,CSC}}$ depends on $B_\mathrm{z}$, since it changes the projection of the spin polarization onto the FM contact magnetization direction. Fitting the Hanle data with a spin diffusion and precession equation \cite{Han2014GrapheneSpintronics,Khokhriakov2020Two-dimensionalGraphene,Vaklinova2016Current-InducedHeterostructures,Hoque2022ChargeNbSe2b} provides a spin lifetime of $\tau_\mathrm{s} \approx \unit[132]{ps}$ and a spin diffusion length of $\lambda_\mathrm{s} \approx \unit[0.74]{\mu m} $ in graphene for spin injection from the TI (for details, see Supplementary Discussion 1). The spin polarization of the TI ($P_{\mathrm{TI}}$) in this device was estimated to be $\unit[1.5]{\%}$ (see Supplementary Discussion 5).  The symmetric Hanle signal suggests that the injected spin into the graphene from the BSTS is polarized along the $y$ axis. This can be further confirmed by in-plane $x$-Hanle measurements, where the magnetic field is applied in-plane along the $x$ direction and lets the injected spin precess along the $yz$ plane. \cite{Benitez2020TunableHeterostructures} The observation of symmetric $x$-Hanle signal along with the corresponding fitting is shown in Figure \ref{fig:1}f. This confirms that the spin that is injected from the BSTS into the graphene channel is polarized in-plane. We could reproducibly observe similar CSC signals with in-plane $y$-spin polarization in Device 2 using spin-switch and Hanle spin precession signals, as presented in Supplementary Discussions 3 and 6.

It should be noted that Device 1 shows a symmetric Hanle signal (Figure \ref{fig:1}e), whereas Device 2 shows an asymmetric Hanle signal (Supplementary Figure 2). The possible mechanism of observing such asymmetric Hanle signals in Device 2 is the charge current distribution inside the BSTS flake having nonzero $x$ and $y$ components (see Supplementary Discussion 3 for details). In Device 1, the observed symmetric spin signal arises due to the charge current having a dominant $x$-axis component.

\subsection*{Geometry and bias dependence of CSC}

\begin{figure}[ht!]
    \centering
    \includegraphics[width=.8\textwidth]{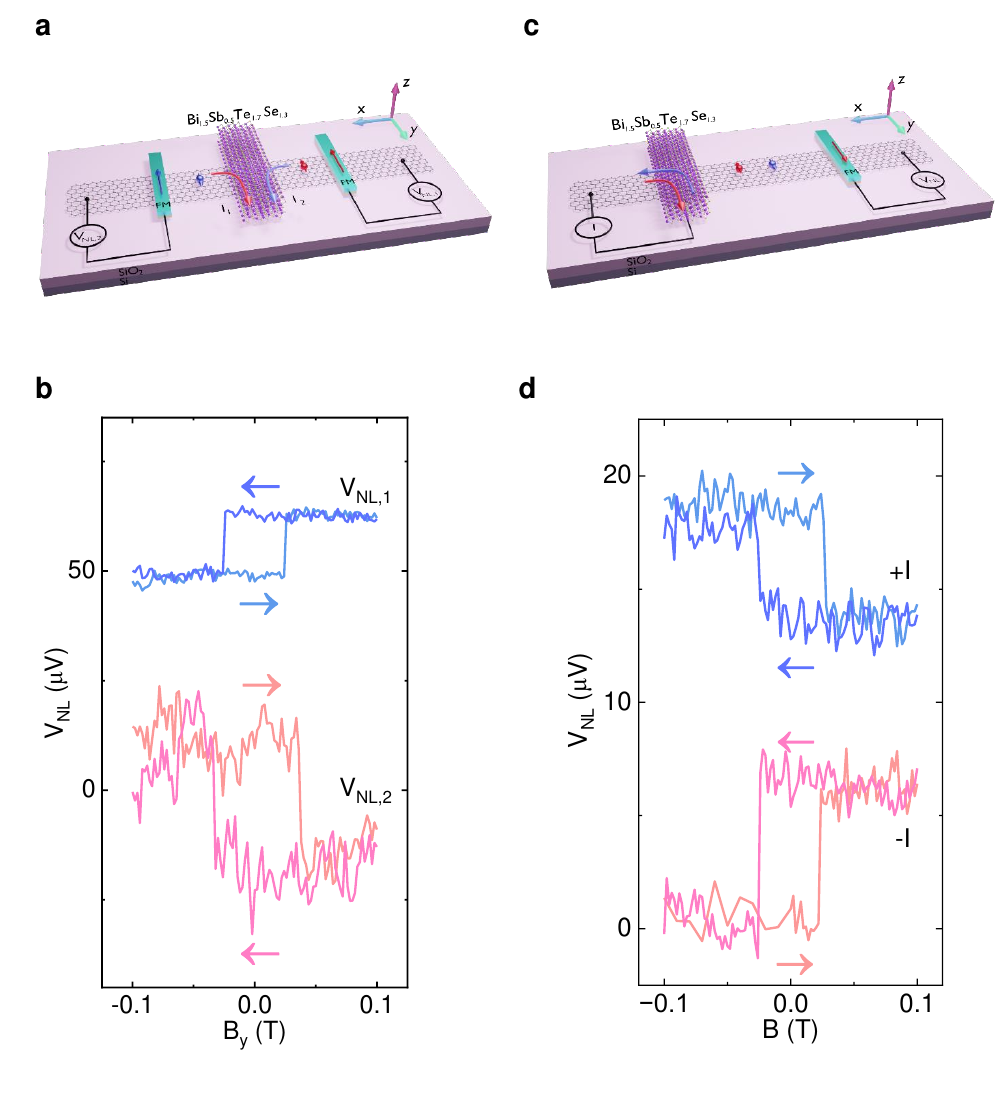}
    \caption{{\footnotesize \textbf{Geometry and bias current dependence of charge-spin conversion signal in graphene/BSTS junction.} \textbf{(a)} Schematic of the device and the two different CSC measurement geometries. \textbf{(b)} Spin-switch signals for each of the two measurement geometries. The measurements were performed at room temperature at $I = \unit[-300]{\mu A}$ and $V_\mathrm{g} = \unit[30]{V}$ for the $V_1$ measurement setup and at $I = \unit[-330]{\mu A}$ and $V_\mathrm{g} = \unit[50]{V}$ for the $V_2$ measurement setup. The channel lengths are $L=\unit[1.84]{\mu m}$ for $V_{\mathrm{NL,1}}$ and $L=\unit[2.7]{\mu m}$ for $V_{\mathrm{NL,2}}$.  \textbf{(c)} Schematic of the device and measurement geometry with arrows indicating the direction of positive and negative bias currents. \textbf{(d)} Spin-switch signals for $I = \unit[-100]{\mu A}$ and $I = \unit[100]{\mu A}$, respectively, at $V_\mathrm{g} = \unit[40]{V}$. The arrows in (b,d) indicate the $B_\mathrm{y}$ field sweep direction. The data in (b,d) was measured in Device 1 and is shifted vertically for clarity.}}
    \label{fig:2}
\end{figure}

In order to further investigate the CSC process in BSTS, experiments were performed for different measurement geometries with spin detection on both sides of the BSTS flake (geometries 1 and 2 in Figure \ref{fig:2}a). In the first geometry, the bias current $I_1$ is applied on the $+x$ side of the BSTS flake and the NL voltage $V_{\mathrm{NL,1}}$ is detected on the $-x$ side of the flake. The second setup uses an opposite geometry with $I_2$ and $V_{\mathrm{NL,2}}$. As shown in Figure \ref{fig:2}b, there is a sign change for the spin-switch signal between the two measurement geometries: setup 1 ($I_1$, $V_{\mathrm{NL,1}}$) gives a high NL voltage when the contact magnetization of the detector is in the $+y$ direction and a low NL voltage for $-y$ contact magnetization, while setup 2 ($I_2$, $V_{\mathrm{NL,2}}$) gives the opposite. The spin-switch signal amplitudes for the two geometries are $\Delta V_{\mathrm{NL,1}} = \unit[12.6]{\mu V}$ and $\Delta V_{\mathrm{NL,2}} = \unit[72.5]{\mu V}$, respectively. It should be noted that differences in amplitude between the two signals can arise from different detector FM contacts being used, which may have different polarizations and/or interface resistances with the graphene. \cite{Kamalakar2014EnhancedNitride} A similar behavior with opposite signs of the spin signals for different measurement geometries is also observed in Device 2 (see Supplementary Discussion 6). The CSC effect must therefore be odd with the $x$ component of the charge current ($I_\mathrm{x}$) because only $I_\mathrm{x}$ changes sign between the two measurement geometries. Considering the conventional CSC processes (SHE, REE and TSML), where charge current ($I$), spin current ($I_\mathrm{s}$) and spin polarization ($s$) need to be orthogonal to each other, this kind of $I_\mathrm{x}$-dependence is expected for in-plane spin along the $y$ axis ($s_\mathrm{y}$) (which we established to be present in our system above). \cite{Sinova2015SpinEffects,Khokhriakov2020Gate-tunableTemperature,Sun2019LargeHeterostructures} These control experiments, however, rule out any unconventional CSC mechanism in the TI, such as the one recently reported in the case of \ce{WTe2} (with lower crystal symmetry), \cite{Zhao2020UnconventionalWTe2} where unconventional CSC was observed to be independent of the $x$ component of the charge current.

The bias current ($I$) dependence of the CSC was examined with an experimental setup as shown in Figure \ref{fig:2}c. The measured spin switch signals (Figure \ref{fig:2}d) show a sign change for the reversal of the charge current direction with similar spin-signal magnitudes ($\Delta V_{\mathrm{NL}}$). We measured $\Delta V_{\mathrm{NL}} = \unit[4.65 \pm 0.7]{\mu V}$ for $I = \unit[100]{\mu A}$ and $\Delta V_{\mathrm{NL}} = \unit[5.8 \pm 0.6]{\mu V}$ for $I = \unit[-100]{\mu A}$. Similar spin signal behavior for opposite $I$ was observed in Device 2 (see Supplementary Discussion 6). Spin signal amplitudes of both spin-switch and Hanle spin precession measurements are shown for several different bias currents in Supplementary Figure 4f. Considering that reversing the bias current means changing between injection and extraction of polarized spin, the observation of a sign change of the spin signal is as expected. \cite{Kovacs-Krausz2020ElectricallyBiTeBr} Additionally, it is also expected that a larger charge current enables a larger spin polarization through CSC.

\subsection*{Gate dependence of CSC signal}

\begin{figure}[ht!]
    \centering
    \includegraphics[width=\textwidth]{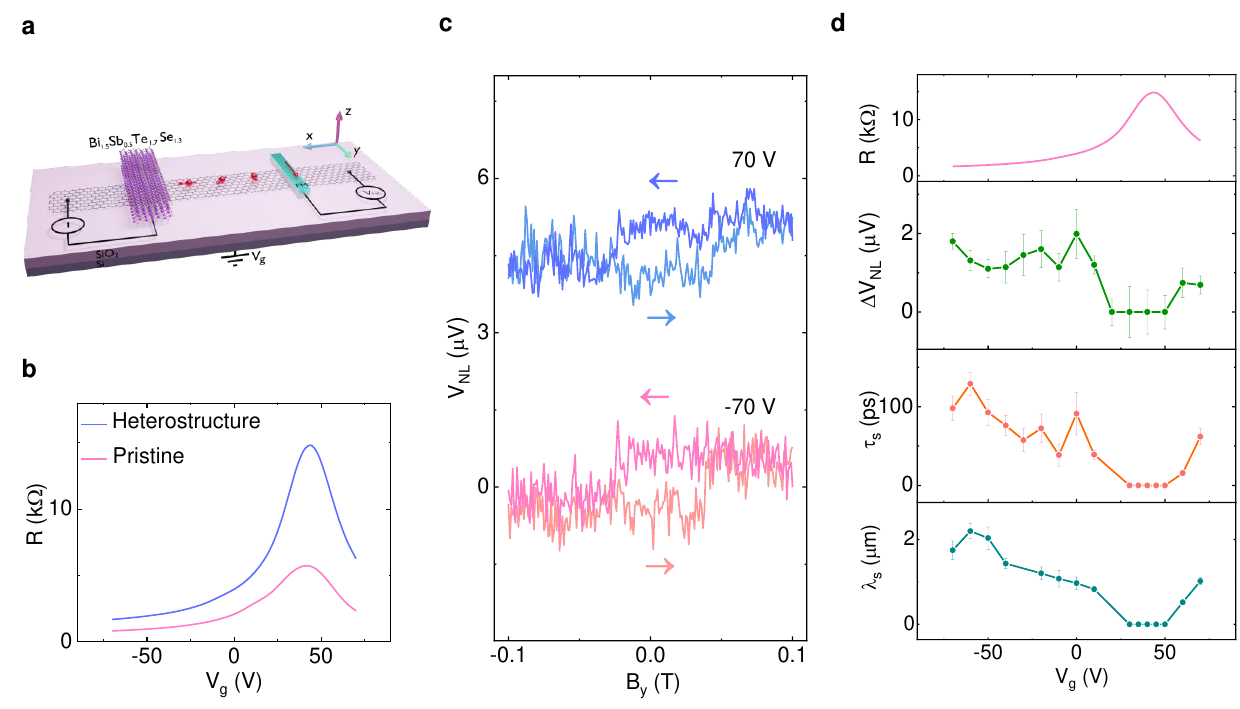}
    \caption{{\footnotesize \textbf{Gate voltage dependence of charge-spin conversion signal in graphene/BSTS heterostructure.} \textbf{(a)} Device geometry for gate-dependent CSC measurements in BSTS with graphene spin-valve device, where a back-gate ($V_\mathrm{g}$) is applied on the Si/\ce{SiO2} substrate. \textbf{(b)} The gate response of channel resistance ($R$) in the heterostructure and pristine graphene regions. \textbf{(c)} Spin-switch signals for electron- and hole-doped graphene regimes measured at $V_\mathrm{g} = \unit[\pm70]{V}$ and $I = \unit[150]{\mu A}$ at room temperature. The data are shifted vertically for clarity. \textbf{(d)} Top panel: The Dirac curve of graphene in the heterostructure region, showing a modulation in resistance $R$ as a function of gate voltage $V_\mathrm{g}$. Three bottom panels: The amplitude of the Hanle signals ($\Delta V_{\mathrm{NL}}$) for the CSC measurements, as well as spin lifetimes $\tau\mathrm{s}$ and spin diffusion lengths $\lambda_\mathrm{s}$ of the graphene, as functions of $V_\mathrm{g}$. The data is shown as mean $\pm$ standard error from fits of the Hanle signals. The measurements were performed at room temperature with $I = \unit[150]{\mu A}$, and the parameters were extracted from fits of the Hanle signals. The data in (b-d) was measured in Device 2.}}
    \label{fig:3}
\end{figure}

The gate dependence of the CSC effect in the TI/graphene heterostructure was investigated in Device 2 in the geometry shown in Figure \ref{fig:3}a, where the Si/\ce{SiO2} substrate was used to apply a back-gate voltage ($V_\mathrm{g}$). First, the gate response ($R$ \emph{vs} $V_\mathrm{g}$) of the graphene channel for pristine graphene and in the graphene/BSTS heterostructure region were measured (Figure \ref{fig:3}b). The Dirac points at $V_\mathrm{g} = \unit[43]{V}$ (for the heterostructure region) and $V_\mathrm{g}=\unit[41]{V}$ (for pristine graphene) indicate a negligible contribution to the doping of graphene from the TI and/or band misalignment between the graphene and BSTS, which could have offset the charge carriers. The difference in resistances comes from the channel across the graphene/BSTS heterostructure being longer ($\unit[6.7]{\mu m}$, compared to $\unit[2.5]{\mu m}$ for the pristine graphene channel). The mobility of the charge carriers was calculated \cite{Lee2020ExtractionModulation} to be $\mu_{\mathrm{HS}} = \unit[1200]{cm^2V^{-1}s^{-1}}$ for the heterostructure graphene and $\mu_{\mathrm{pr}} = \unit[1000]{cm^2V^{-1}s^{-1}}$ for the pristine graphene, as expected for CVD graphene. \cite{Khokhriakov2020Gate-tunableTemperature,Hoque2021All-electricalTemperature,Khokhriakov2018TailoringHeterostructures,Hoque2020Charge-spinHeterostructures,Hoque2023Spin-valleyHeterostructure}

The gate-dependent spin-switch measurements of the CSC effect are presented in Figure \ref{fig:3}c for p- ($V_\mathrm{g} = \unit[-70]{V}$) and n-doped ($V_\mathrm{g} = \unit[70]{V}$) regimes of the graphene channel. It is found that the sign of the signal remains the same both for the electron- and the hole-doped regimes. As the TI is not very sensitive to gate voltage in our experiments (because of shielding by the graphene and the thickness of the TI flake itself, see Supplementary Discussion 7), the modulation of the Fermi level in graphene is expected to cause a negligible Fermi level tuning of BSTS. The absence of a sign change between the CSC signals for p- and n-doped graphene regions further confirms the spin origin of the signal and rules out any artifact arising from the stray magnetic field from the FM detector contact on the graphene. \cite{Karpiak20181DHeterostructures}

Detailed gate-dependence of the spin signals and spin transport parameters are shown in Figure \ref{fig:3}d. The top panel in Figure \ref{fig:3}d presents the channel resistance of the graphene in the heterostructure region. The amplitude of the Hanle signals ($\Delta V_{\mathrm{NL}}$) shows a minimum around the Dirac point of graphene. The gate dependence of the spin lifetime ($\tau_\mathrm{s}$) and the spin diffusion length ($\lambda_\mathrm{s}$) of the graphene after the spin is injected from the TI are depicted in the two bottom panels of Figure \ref{fig:3}d. A weak linear dependency of $\tau_\mathrm{s}$ and $\lambda_\mathrm{s}$ with $V_\mathrm{g}$ can be seen farther away from the Dirac point of the graphene. The disappearance of the spin signal around the graphene Dirac point is likely due to a conductivity mismatch between the graphene channel and the TI and/or the FM contact. The resistance of the graphene is at a maximum at the Dirac point, which increases the likelihood of injected spins diffusing back or being reabsorbed by the TI instead of propagating along the graphene channel, leading to a decreased spin injection and detection efficiency and, subsequently, to a low spin signal. \cite{Han2010TunnelingGraphene,Kamalakar2015LongGraphene,Kamalakar2014EnhancedNitride} A further discussion about the gate response of the CSC effect in TI/graphene heterostructure and of the spin transport in graphene, along with Hanle measurements, is provided in Supplementary Discussions 8-9.

\subsection*{Inverse CSC effects}

\begin{figure}[ht!]
    \centering
    \includegraphics[width=\textwidth]{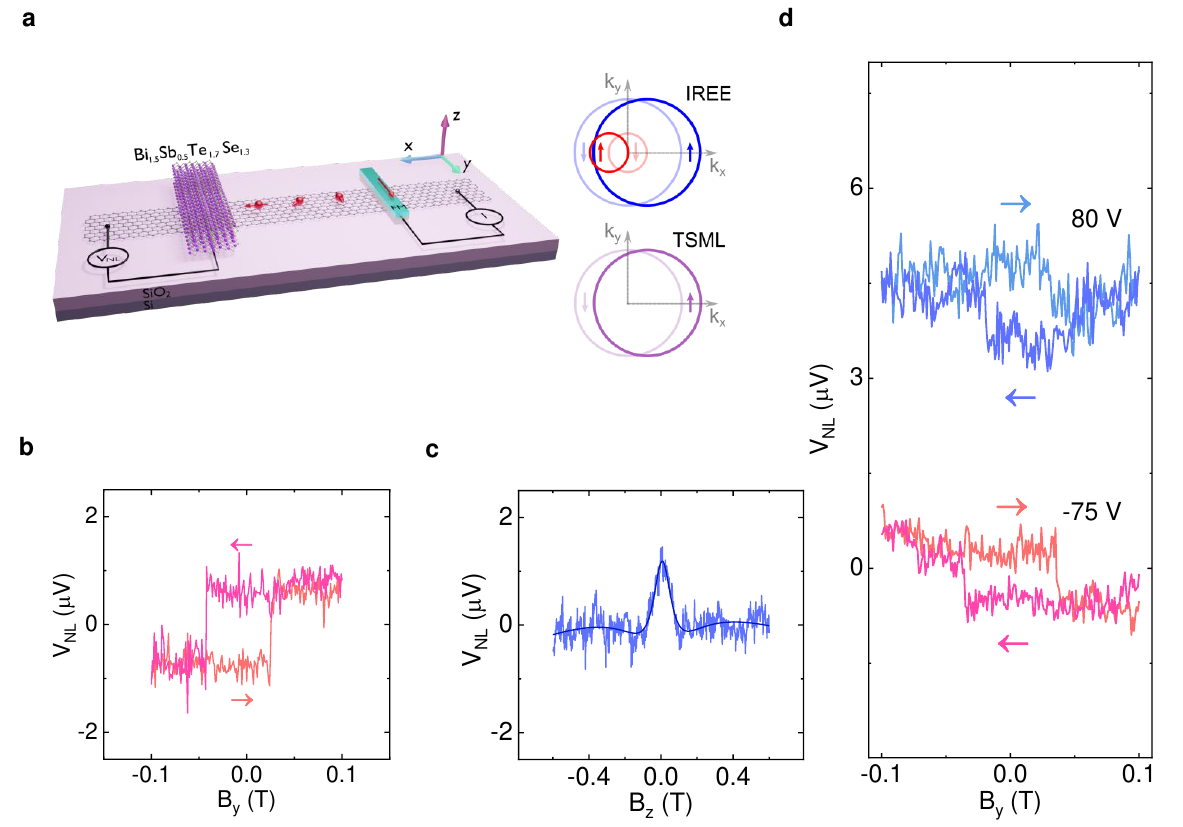}
    \caption{{\footnotesize \textbf{The inverse charge-spin conversion signal in graphene/BSTS heterostructure.} \textbf{(a)} Schematic of the device and electrical connections for measuring ICSC in TI, where spin is injected from a FM contact into graphene and later detected by the TI in a NL geometry. Inset: Schematic of non-equilibrium Fermi contours for ICSC through the inverse Rashba-Edelstein effect (IREE) and TSML. \textbf{(b)} NL spin-switch signal for ICSC measurement with $I = \unit[-150]{\mu A}$ and $V_\mathrm{g} = \unit[-70]{V}$. \textbf{(c)} Hanle spin-precession signal along with Hanle fitting for ICSC measurement at $I = \unit[-200]{\mu A}$ and $V_\mathrm{g} = \unit[-75]{V}$. A spin lifetime $\tau_\mathrm{s} \approx \unit[75]{ps}$ and a spin diffusion length $\lambda_\mathrm{s} \approx \unit[1.3]{\mu m}$ were extracted from the fitting. \textbf{(d)}  NL spin-switch signals for ICSC setup at $V_\mathrm{g} = \unit[80]{V}$ and $V_\mathrm{g} = \unit[-75]{V}$, respectively, with $I = \unit[-100]{\mu A}$. The data in (b-d) was measured in Device 2.}}
    \label{fig:4}
\end{figure}

Inverse CSC (ICSC) experiments were performed in the TI to verify the Onsager reciprocity in the system and to detect spin polarization from the graphene channel with the TI. The spin current is injected from the FM contact into the graphene, and is then absorbed by the TI and produces a NL voltage due to the ICSC effect, as illustrated in Figure \ref{fig:4}a. The spin polarization causes the Fermi surface spin textures of the TI to shift to a non-equilibrium position in the way shown in the inset of Figure \ref{fig:4}a, which is expected to give rise to a charge current. Spin-switch measurements indeed showed ICSC  effects (Figure \ref{fig:4}b) at room temperature. It can clearly be seen that the CSC (bottom signal of Supplementary Figure 4d) and the ICSC (Figure \ref{fig:4}b) obey reciprocity by generating spin signals of opposite sign for similar bias current and the same geometry, due to the spin current having opposite directions in the BSTS flake for the two cases. Hanle spin precession measurements for the ICSC measurement geometry (see Figure \ref{fig:4}c) are used to verify the spin origin of the observed ICSC signal. In the ICSC measurements, the extracted spin lifetime and spin diffusion length in graphene are $\tau_\mathrm{s} \approx \unit[75]{ps}$ and $\lambda_\mathrm{s} \approx \unit[1.3]{\mu m} $, respectively. These values are very close to the parameters obtained from the direct CSC measurements.

The spin polarization of BSTS in the ICSC measurement was determined to about $\unit[0.2]{\%}$ in Device 2, slightly higher than for the CSC counterpart ($\unit[0.1]{\%}$). The reason for this slight difference could be due to different spin injection and detection efficiencies of the graphene/BSTS and the graphene/FM contacts with and without bias voltages. Interestingly, the Hanle spin precession signals from Device 2 are asymmetric in the CSC measurements (Supplementary Figure 2), whereas the signals have a dominating symmetric component in the ICSC measurement geometry (Supplementary Figures 7e-f). This discrepancy can be understood by considering the electric field that is applied across the graphene/BSTS interface in the CSC measurements, which gives rise to the current distribution in the BSTS flake with nonzero $x$, $y$ and $z$ components, and is different from the ICSC measurements.

ICSC measurements in two geometries with opposite $x$ components of the spin current showed a change in the NL voltage sign when the polarization of the injected spin current was reversed (see Supplementary Discussion 10). This means that the ICSC has a dependence on the $x$ component of the spin current (${I_\mathrm{s}}_\mathrm{x}$), similar to the $I_\mathrm{x}$-dependence of the direct CSC. Next, ICSC signals for both n- and p-doped graphene regimes are shown in Figure \ref{fig:4}d and the sign of the spin-switch signals remains the same, similar to the direct CSC measurements (Figure \ref{fig:3}c).

The spin Hall angle in the TI is estimated \cite{Yan2017LargeHeterostructure,Safeer2019LargeTemperature,Hoque2020Charge-spinHeterostructures} to be $\theta_{\mathrm{SHE}} \geq \unit[2.8]{\%}$ and the spin Hall length scale is subsequently calculated \cite{Safeer2019LargeTemperature,Hoque2020Charge-spinHeterostructures} as $\lambda_{\mathrm{SHE}} = \theta_{\mathrm{SHE}} \lambda_\mathrm{s}^{\mathrm{BSTS}} \geq \unit[0.55]{nm}$, where $\lambda_\mathrm{s}^{\mathrm{BSTS}}$ is the spin diffusion length of the BSTS. Similarly, the REE efficiency parameter is estimated \cite{Khokhriakov2020Gate-tunableTemperature} to be $\alpha_{\mathrm{REE}} = \unit[2.8]{\%}$ and the Rashba-Edelstein length scale is then calculated \cite{Khokhriakov2020Gate-tunableTemperature} as $\lambda_{\mathrm{REE}} \leq \alpha_{\mathrm{REE}} \lambda_\mathrm{s} = \unit[17]{nm}$, where the spin diffusion length of graphene, $\lambda_\mathrm{s}$, is chosen as an upper bound of the heterostructure spin diffusion length. However, these values are only to be regarded as rough estimates for order-of-magnitude-level comparisons, because the adapted models are not accurate with the used measurement geometries (see discussion in Supplementary Discussion 12).

\subsection*{Possible origins of the (I)CSC effects}

Here, we discuss the possible origins of the (I)CSC effects in the TI. First and foremost, the spin-switch and Hanle spin precession measurements unequivocally prove the spin origin of the observed (I)CSC signals. Next, the observation of a sign change of the spin signals with reversing the bias current rules out thermal effects as the origin of the observed spin signals. \cite{Sierra2018ThermoelectricGraphene} Consequently, the observed spin signal can originate either from CSC in the TI or from proximity-induced CSC of the graphene in the TI/graphene heterostructure. In our measurements, we can rule out the proximity-induced SHE in graphene, as it should generate an out-of-plane spin polarization with an anti-symmetric $x$-Hanle signal. \cite{Safeer2019Room-TemperatureHeterostructures,Ingla-aynes2022OmnidirectionalHeterostructuresb,Rodriguez-Vega2017GiantHeterostructures} Furthermore, for proximity-induced CSC (both SHE and REE), the spin signals should have opposite signs for electron- and hole-doped regimes,  which is not observed in our measurements \cite{Khokhriakov2020Gate-tunableTemperature, Li2020Gate-TunableTemperature,Benitez2020TunableHeterostructures,Song2018SpinHeterostructures}. This indicates that the CSC does not originate from proximitized graphene but only from CSC in the TI itself in our device.

The origin of the CSC signal can be either of the SHE, the REE and the TSML of the TI at room temperature, or from a combination of the three effects. It is, however, difficult to make a definitive distinction between these three effects in our system. As has previously been shown, \cite{Dankert2018OriginInsulators} the contribution from the TSML of the TSS in BSTS is only dominant below $\unit[100]{K}$ (although it may still have a nonzero contribution at room temperature). Furthermore, the SHE, the REE and the TSML have similar geometric dependencies, as mentioned above, where the charge current, spin current and spin polarization have to be orthogonal to each other. The SHE and the REE can, in principle, give a sign change response for applied gate voltages, but this is not expected in our system, since this would necessitate the Fermi level to be tuned across the TI band gap and would therefore require a very large applied gate voltage in a heavily doped TI such as BSTS. \cite{Khokhriakov2020Gate-tunableTemperature,Hoque2021All-electricalTemperature} The TSML should not give a sign change either, even when the Fermi level is tuned across the TI Dirac point, due to the opposite chirality of the TSML above and below the Dirac point, which counteracts the effect of the changed charge carrier types. \cite{Dankert2018OriginInsulators,Khokhriakov2020Gate-tunableTemperature}

Because of the similarities between the SHE, the REE and the TSML of TI, it is not possible to further discern the exact contributions of the CSC in BSTS at room temperature from our measurements. Due to the Onsager reciprocity, the above arguments hold also for the ICSC, and the same three effects are identified as its possible origins. Future temperature- and TI thickness-dependent measurements should provide indication of bulk (SHE and bulk-state REE) and surface (REE of trivial surface states and TSML) (I)CSC contributions of TIs.

The (I)CSC effects as well as spin injection and detection in the graphene channel are sensitive to $I$, $V_\mathrm{g}$, the graphene/TI interface condition, and the spin transport properties in the graphene channel. When changing $V_\mathrm{g}$ or measurement geometry, the spin transport parameters are also likely to change. In addition to this, the quality of contact interfaces are likely to change somewhat over time, as a result of the applied bias currents. Because of this, it is often necessary to optimize the applied $I$ and $V_\mathrm{g}$ accordingly in order to observe clear spin signals to get the best transport parameters.

In summary, we have demonstrated room temperature CSC and its inverse effect (ICSC) in the TI BSTS, using a hybrid device with a graphene spin-valve. The lateral nonlocal spin-switch and Hanle spin-precession measurements, supported by the detailed dependence on bias current, gate voltage, different geometries and magnetization orientations, prove the CSC effects in BSTS. We could conclude that both the CSC and the ICSC originate from the SHE or REE, from the TSML of the TI surface states, or from a combination of these effects at room temperature. Our results prove CSC effects in TIs at room temperature, which is considered a potential candidate for SOT-based memory, logic and neuromorphic computing technologies. \cite{Dieny2020OpportunitiesIndustry,Sierra2021VanOpto-spintronics} With the recent progress in graphene spintronics, \cite{Khokhriakov2022MultifunctionalCircuits,Khokhriakov2020Two-dimensionalGraphene,Khokhriakov2020RobustBoundaries} the attainment of TIs as spin injectors/detectors can also provide substantial advances in all-electrical spin-based devices and integrated circuits in 2D architectures.


\section*{Methods}
\subsection*{Fabrication}
Patterned graphene stripes were processed from $\unit[7]{mm}$ chips cut from a $\unit[4]{''}$ wafer with CVD graphene (from Grolltex Inc.) through electron beam lithography and oxygen plasma etching. The \ce{Bi_{1.5}Sb_{0.5}Te_{1.7}Se_{1.3}} (BSTS) flakes were exfoliated mechanically on top of the graphene stripes inside a glovebox in \ce{N2} atmosphere. Nonmagnetic reference contacts ($\unit[20]{nm}$ \ce{Cr}/$\unit[90]{nm}$ \ce{Au}) on BSTS and graphene were made using electron beam lithography, thin film deposition by electron beam evaporation and liftoff in acetone at $\unit[65]{\degree C}$. Ferromagnetic tunnel contacts ($\unit[\sim 1]{nm}$ \ce{TiO2}/$\unit[90]{nm}$ \ce{Co}) were subsequently made similarly using electron beam lithography, thin film deposition and liftoff processes. The \ce{TiO2} tunnel barriers were prepared by electron beam evaporation of \ce{Ti} in two steps, each followed by \emph{in-situ} oxidation in a pure oxygen atmosphere.

\subsection*{Measurements}
The bias current and gate voltage were applied using a Keithley 6221 current source and a Keithley 2400 source meter, respectively, and the nonlocal voltage was detected by a Keithley 2182A nanovoltmeter.

\section*{Data Availability}
The data supporting the findings of this study are available from the corresponding author upon reasonable request.

\section*{Acknowledgements}
The authors acknowledge financial supports from EU Graphene Flagship (Core 3, No. 881603), Swedish Research Council VR project grants (No. 2021–04821), 2D TECH VINNOVA competence center (No. 2019-00068), FLAG-ERA project 2DSOTECH (VR No. 2021-05925), Graphene center, EI Nano, and AoA Materials program at Chalmers University of Technology. We acknowledge the help of staff at Quantum Device Physics and Nanofabrication laboratory in our department at Chalmers. Devices were fabricated at Nanofabrication laboratory, Myfab, MC2, Chalmers.

\section*{Author Contributions}
A.M.H. and L.S. equally contributed to this manuscript. A.M.H. and L.S. fabricated and characterized the devices. D.K., B.Z., S.P.D. participated in discussions for device preparation and measurements. A.M.H., L.S. and S.P.D. conceived the idea and designed the experiments. A.M.H., L.S., and S.P.D. analyzed and interpreted the experimental data, compiled the figures, and wrote the manuscript with inputs from all co-authors. S.P.D. supervised the research project. Some of these results are part of Lars Sjöström’s master’s thesis "Room Temperature Charge-Spin Interconversion in a Topological Insulator and Graphene Heterostructure", Chalmers Digital Repository (2021).

\section*{Competing Interests}
The authors declare no competing financial or non-financial interests.


\bibliography{references_abbr}

\begin{thebibliography}{10}
\expandafter\ifx\csname url\endcsname\relax
  \def\url#1{\burl{#1}}\fi
\expandafter\ifx\csname urlprefix\endcsname\relax\def\urlprefix{URL }\fi
\providecommand{\bibinfo}[2]{#2}
\providecommand{\eprint}[2][]{\url{#2}}
\providecommand{\doi}[1]{\url{https://doi.org/#1}}
\bibcommenthead

\bibitem{Tokura2017EmergentMaterials}
\bibinfo{author}{Tokura, Y.}, \bibinfo{author}{Kawasaki, M.} \& \bibinfo{author}{Nagaosa, N.}
\newblock \bibinfo{title}{{Emergent functions of quantum materials}}.
\newblock \emph{\bibinfo{journal}{Nat. Phys.}} \textbf{\bibinfo{volume}{13}}, \bibinfo{pages}{1056--1068} (\bibinfo{year}{2017}).

\bibitem{Qi2011TopologicalSuperconductors}
\bibinfo{author}{Qi, X.-L.} \& \bibinfo{author}{Zhang, S.-C.}
\newblock \bibinfo{title}{{Topological insulators and superconductors}}.
\newblock \emph{\bibinfo{journal}{Rev. Mod. Phys.}} \textbf{\bibinfo{volume}{83}}, \bibinfo{pages}{1057--1110} (\bibinfo{year}{2011}).

\bibitem{Sierra2021VanOpto-spintronics}
\bibinfo{author}{Sierra, J.~F.}, \bibinfo{author}{Fabian, J.}, \bibinfo{author}{Kawakami, R.~K.}, \bibinfo{author}{Roche, S.} \& \bibinfo{author}{Valenzuela, S.~O.}
\newblock \bibinfo{title}{{Van der Waals heterostructures for spintronics and opto-spintronics}}.
\newblock \emph{\bibinfo{journal}{Nat. Nanotechnol.}} \textbf{\bibinfo{volume}{16}}, \bibinfo{pages}{856--868} (\bibinfo{year}{2021}).

\bibitem{Hasan2010Colloquium:Insulators}
\bibinfo{author}{Hasan, M.~Z.} \& \bibinfo{author}{Kane, C.~L.}
\newblock \bibinfo{title}{{Colloquium: Topological insulators}}.
\newblock \emph{\bibinfo{journal}{Rev. Mod. Phys.}} \textbf{\bibinfo{volume}{82}}, \bibinfo{pages}{3045--3067} (\bibinfo{year}{2010}).

\bibitem{Xia2009ObservationSurface}
\bibinfo{author}{Xia, Y.} \emph{et~al.}
\newblock \bibinfo{title}{{Observation of a large-gap topological-insulator class with a single Dirac cone on the surface}}.
\newblock \emph{\bibinfo{journal}{Nat. Phys.}} \textbf{\bibinfo{volume}{5}}, \bibinfo{pages}{398--402} (\bibinfo{year}{2009}).

\bibitem{Hsieh2009ARegime}
\bibinfo{author}{Hsieh, D.} \emph{et~al.}
\newblock \bibinfo{title}{{A tunable topological insulator in the spin helical Dirac transport regime}}.
\newblock \emph{\bibinfo{journal}{Nature}} \textbf{\bibinfo{volume}{460}}, \bibinfo{pages}{1101--1105} (\bibinfo{year}{2009}).

\bibitem{Lei2022GrapheneDevices}
\bibinfo{author}{Lei, Y.} \emph{et~al.}
\newblock \bibinfo{title}{{Graphene and Beyond: Recent Advances in Two-Dimensional Materials Synthesis, Properties, and Devices}}.
\newblock \emph{\bibinfo{journal}{ACS Nanosci.}} \textbf{\bibinfo{volume}{2}}, \bibinfo{pages}{450--485} (\bibinfo{year}{2022}).

\bibitem{Chakraborty2022ChallengesDevices}
\bibinfo{author}{Chakraborty, S.~K.}, \bibinfo{author}{Kundu, B.}, \bibinfo{author}{Nayak, B.}, \bibinfo{author}{Dash, S.~P.} \& \bibinfo{author}{Sahoo, P.~K.}
\newblock \bibinfo{title}{{Challenges and opportunities in 2D heterostructures for electronic and optoelectronic devices}}.
\newblock \emph{\bibinfo{journal}{iScience}} \textbf{\bibinfo{volume}{25}}, \bibinfo{pages}{103942} (\bibinfo{year}{2022}).

\bibitem{Wu2021MagneticInsulators}
\bibinfo{author}{Wu, H.} \emph{et~al.}
\newblock \bibinfo{title}{{Magnetic memory driven by topological insulators}}.
\newblock \emph{\bibinfo{journal}{Nat. Commun.}} \textbf{\bibinfo{volume}{12}}, \bibinfo{pages}{6251} (\bibinfo{year}{2021}).

\bibitem{Mellnik2014Spin-transferInsulator}
\bibinfo{author}{Mellnik, A.~R.} \emph{et~al.}
\newblock \bibinfo{title}{{Spin-transfer torque generated by a topological insulator}}.
\newblock \emph{\bibinfo{journal}{Nature}} \textbf{\bibinfo{volume}{511}}, \bibinfo{pages}{449--451} (\bibinfo{year}{2014}).

\bibitem{Manchon2019Current-inducedSystems}
\bibinfo{author}{Manchon, A.} \emph{et~al.}
\newblock \bibinfo{title}{{Current-induced spin-orbit torques in ferromagnetic and antiferromagnetic systems}}.
\newblock \emph{\bibinfo{journal}{Rev. Mod. Phys.}} \textbf{\bibinfo{volume}{91}}, \bibinfo{pages}{035004} (\bibinfo{year}{2019}).

\bibitem{DC2023ObservationMnPd3}
\bibinfo{author}{DC, M.} \emph{et~al.}
\newblock \bibinfo{title}{{Observation of anti-damping spin-orbit torques generated by in-plane and out-of-plane spin polarizations in MnPd3}}.
\newblock \emph{\bibinfo{journal}{Nat. Mater.}} \textbf{\bibinfo{volume}{22}}, \bibinfo{pages}{591--598} (\bibinfo{year}{2023}).

\bibitem{Fan2014MagnetizationHeterostructure}
\bibinfo{author}{Fan, Y.} \emph{et~al.}
\newblock \bibinfo{title}{{Magnetization switching through giant spin-orbit torque in a magnetically doped topological insulator heterostructure}}.
\newblock \emph{\bibinfo{journal}{Nat. Mater.}} \textbf{\bibinfo{volume}{13}}, \bibinfo{pages}{699--704} (\bibinfo{year}{2014}).

\bibitem{Khang2018ASwitching}
\bibinfo{author}{Khang, N. H.~D.}, \bibinfo{author}{Ueda, Y.} \& \bibinfo{author}{Hai, P.~N.}
\newblock \bibinfo{title}{{A conductive topological insulator with large spin Hall effect for ultralow power spin-orbit torque switching}}.
\newblock \emph{\bibinfo{journal}{Nat. Mater.}} \textbf{\bibinfo{volume}{17}}, \bibinfo{pages}{808--813} (\bibinfo{year}{2018}).

\bibitem{Han2017Room-TemperatureInsulator}
\bibinfo{author}{Han, J.} \emph{et~al.}
\newblock \bibinfo{title}{{Room-Temperature Spin-Orbit Torque Switching Induced by a Topological Insulator}}.
\newblock \emph{\bibinfo{journal}{Phys. Rev. Lett.}} \textbf{\bibinfo{volume}{119}}, \bibinfo{pages}{077702} (\bibinfo{year}{2017}).

\bibitem{Valla2012PhotoemissionInsulator}
\bibinfo{author}{Valla, T.}, \bibinfo{author}{Pan, Z.~H.}, \bibinfo{author}{Gardner, D.}, \bibinfo{author}{Lee, Y.~S.} \& \bibinfo{author}{Chu, S.}
\newblock \bibinfo{title}{{Photoemission spectroscopy of magnetic and nonmagnetic impurities on the surface of the Bi 2Se 3 topological insulator}}.
\newblock \emph{\bibinfo{journal}{Phys. Rev. Lett.}} \textbf{\bibinfo{volume}{108}}, \bibinfo{pages}{117601} (\bibinfo{year}{2012}).

\bibitem{Bonell2020ControlHeterostructures}
\bibinfo{author}{Bonell, F.} \emph{et~al.}
\newblock \bibinfo{title}{{Control of Spin-Orbit Torques by Interface Engineering in Topological Insulator Heterostructures}}.
\newblock \emph{\bibinfo{journal}{Nano Lett.}} \textbf{\bibinfo{volume}{20}}, \bibinfo{pages}{5893--5899} (\bibinfo{year}{2020}).

\bibitem{Li2012MagneticBi2Se3}
\bibinfo{author}{Li, J.} \emph{et~al.}
\newblock \bibinfo{title}{{Magnetic dead layer at the interface between a Co film and the topological insulator Bi2Se3}}.
\newblock \emph{\bibinfo{journal}{Phys. Rev. B: Condens. Matter.}} \textbf{\bibinfo{volume}{86}}, \bibinfo{pages}{054430} (\bibinfo{year}{2012}).

\bibitem{Li2014ElectricalBi2Se3}
\bibinfo{author}{Li, C.~H.} \emph{et~al.}
\newblock \bibinfo{title}{{Electrical detection of charge-current-induced spin polarization due to spin-momentum locking in Bi2Se3}}.
\newblock \emph{\bibinfo{journal}{Nat. Nanotechnol.}} \textbf{\bibinfo{volume}{9}}, \bibinfo{pages}{218--224} (\bibinfo{year}{2014}).

\bibitem{Dankert2015RoomInsulators}
\bibinfo{author}{Dankert, A.}, \bibinfo{author}{Geurs, J.}, \bibinfo{author}{Kamalakar, M.~V.}, \bibinfo{author}{Charpentier, S.} \& \bibinfo{author}{Dash, S.~P.}
\newblock \bibinfo{title}{{Room temperature electrical detection of spin polarized currents in topological insulators}}.
\newblock \emph{\bibinfo{journal}{Nano Lett.}} \textbf{\bibinfo{volume}{15}}, \bibinfo{pages}{7976--7981} (\bibinfo{year}{2015}).

\bibitem{Ando2014ElectricalBi1.5Sb0.5Te1.7Se1.3}
\bibinfo{author}{Ando, Y.} \emph{et~al.}
\newblock \bibinfo{title}{{Electrical Detection of the Spin Polarization Due to Charge Flow in the Surface State of the Topological Insulator Bi1.5Sb0.5Te1.7Se1.3}}.
\newblock \emph{\bibinfo{journal}{Nano Lett.}} \textbf{\bibinfo{volume}{14}}, \bibinfo{pages}{6226--6230} (\bibinfo{year}{2014}).

\bibitem{Tang2014ElectricalInsulator}
\bibinfo{author}{Tang, J.} \emph{et~al.}
\newblock \bibinfo{title}{{Electrical Detection of Spin-Polarized Conduction in (Bi0.53Sb0.47)2Te3 Topological Insulator}}.
\newblock \emph{\bibinfo{journal}{Nano Lett.}} \textbf{\bibinfo{volume}{14}}, \bibinfo{pages}{5423--5429} (\bibinfo{year}{2014}).

\bibitem{Liu2015Spin-polarizedInsulators}
\bibinfo{author}{Liu, L.} \emph{et~al.}
\newblock \bibinfo{title}{{Spin-polarized tunneling study of spin-momentum locking in topological insulators}}.
\newblock \emph{\bibinfo{journal}{Phys. Rev. B: Condens. Matter.}} \textbf{\bibinfo{volume}{91}}, \bibinfo{pages}{235437} (\bibinfo{year}{2015}).

\bibitem{Tian2015ElectricalBi2Te2Se}
\bibinfo{author}{Tian, J.}, \bibinfo{author}{Miotkowski, I.}, \bibinfo{author}{Hong, S.} \& \bibinfo{author}{Chen, Y.~P.}
\newblock \bibinfo{title}{{Electrical injection and detection of spin-polarized currents in topological insulator Bi2Te2Se}}.
\newblock \emph{\bibinfo{journal}{Sci. Rep.}} \textbf{\bibinfo{volume}{5}}, \bibinfo{pages}{14293} (\bibinfo{year}{2015}).

\bibitem{Li2016DirectStates}
\bibinfo{author}{Li, C.~H.}, \bibinfo{author}{Vant~Erve, O.~M.}, \bibinfo{author}{Rajput, S.}, \bibinfo{author}{Li, L.} \& \bibinfo{author}{Jonker, B.~T.}
\newblock \bibinfo{title}{{Direct comparison of current-induced spin polarization in topological insulator Bi2Se3 and InAs Rashba states}}.
\newblock \emph{\bibinfo{journal}{Nat. Commun.}} \textbf{\bibinfo{volume}{7}}, \bibinfo{pages}{13518} (\bibinfo{year}{2016}).

\bibitem{Tian2017ObservationBattery}
\bibinfo{author}{Tian, J.}, \bibinfo{author}{Hong, S.}, \bibinfo{author}{Miotkowski, I.}, \bibinfo{author}{Datta, S.} \& \bibinfo{author}{Chen, Y.~P.}
\newblock \bibinfo{title}{{Observation of current-induced, long-lived persistent spin polarization in a topological insulator: A rechargeable spin battery}}.
\newblock \emph{\bibinfo{journal}{Sci. Adv.}} \textbf{\bibinfo{volume}{3}}, \bibinfo{pages}{e1602531} (\bibinfo{year}{2017}).

\bibitem{Dankert2018OriginInsulators}
\bibinfo{author}{Dankert, A.} \emph{et~al.}
\newblock \bibinfo{title}{{Origin and evolution of surface spin current in topological insulators}}.
\newblock \emph{\bibinfo{journal}{Phys. Rev. B: Condens. Matter.}} \textbf{\bibinfo{volume}{97}}, \bibinfo{pages}{125414} (\bibinfo{year}{2018}).

\bibitem{Han2014GrapheneSpintronics}
\bibinfo{author}{Han, W.}, \bibinfo{author}{Kawakami, R.~K.}, \bibinfo{author}{Gmitra, M.} \& \bibinfo{author}{Fabian, J.}
\newblock \bibinfo{title}{{Graphene spintronics}}.
\newblock \emph{\bibinfo{journal}{Nat. Nanotechnol.}} \textbf{\bibinfo{volume}{9}}, \bibinfo{pages}{794--807} (\bibinfo{year}{2014}).

\bibitem{Avsar2020Colloquium:Materials}
\bibinfo{author}{Avsar, A.} \emph{et~al.}
\newblock \bibinfo{title}{{Colloquium: Spintronics in graphene and other two-dimensional materials}}.
\newblock \emph{\bibinfo{journal}{Rev. Mod. Phys.}} \textbf{\bibinfo{volume}{92}}, \bibinfo{pages}{021003} (\bibinfo{year}{2020}).

\bibitem{Safeer2022ReliabilityValves}
\bibinfo{author}{Safeer, C.~K.} \emph{et~al.}
\newblock \bibinfo{title}{{Reliability of spin-to-charge conversion measurements in graphene-based lateral spin valves}}.
\newblock \emph{\bibinfo{journal}{2D Mater.}} \textbf{\bibinfo{volume}{9}}, \bibinfo{pages}{015024} (\bibinfo{year}{2022}).

\bibitem{Khokhriakov2020Two-dimensionalGraphene}
\bibinfo{author}{Khokhriakov, D.}, \bibinfo{author}{Karpiak, B.}, \bibinfo{author}{Hoque, A.~M.} \& \bibinfo{author}{Dash, S.~P.}
\newblock \bibinfo{title}{{Two-dimensional spintronic circuit architectures on large scale graphene}}.
\newblock \emph{\bibinfo{journal}{Carbon}} \textbf{\bibinfo{volume}{161}}, \bibinfo{pages}{892--899} (\bibinfo{year}{2020}).

\bibitem{Khokhriakov2020RobustBoundaries}
\bibinfo{author}{Khokhriakov, D.} \emph{et~al.}
\newblock \bibinfo{title}{{Robust spin interconnect with isotropic spin dynamics in chemical vapor deposited graphene layers and boundaries}}.
\newblock \emph{\bibinfo{journal}{ACS Nano}} \textbf{\bibinfo{volume}{14}}, \bibinfo{pages}{15864--15873} (\bibinfo{year}{2020}).

\bibitem{Khokhriakov2022MultifunctionalCircuits}
\bibinfo{author}{Khokhriakov, D.} \emph{et~al.}
\newblock \bibinfo{title}{{Multifunctional Spin Logic Operations in Graphene Spin Circuits}}.
\newblock \emph{\bibinfo{journal}{Phys. Rev. Appl.}} \textbf{\bibinfo{volume}{18}}, \bibinfo{pages}{064063} (\bibinfo{year}{2022}).

\bibitem{Vaklinova2016Current-InducedHeterostructures}
\bibinfo{author}{Vaklinova, K.}, \bibinfo{author}{Hoyer, A.}, \bibinfo{author}{Burghard, M.} \& \bibinfo{author}{Kern, K.}
\newblock \bibinfo{title}{{Current-Induced Spin Polarization in Topological Insulator-Graphene Heterostructures}}.
\newblock \emph{\bibinfo{journal}{Nano Lett.}} \textbf{\bibinfo{volume}{16}}, \bibinfo{pages}{2595--2602} (\bibinfo{year}{2016}).

\bibitem{Voerman2019Spin-MomentumMeasurements}
\bibinfo{author}{Voerman, J.~A.}, \bibinfo{author}{Li, C.}, \bibinfo{author}{Huang, Y.} \& \bibinfo{author}{Brinkman, A.}
\newblock \bibinfo{title}{{Spin-Momentum Locking in the Gate Tunable Topological Insulator BiSbTeSe2 in Non-Local Transport Measurements}}.
\newblock \emph{\bibinfo{journal}{Adv. Electron. Mater.}} \textbf{\bibinfo{volume}{5}}, \bibinfo{pages}{1900334} (\bibinfo{year}{2019}).

\bibitem{Zhao2023AHeterostructure}
\bibinfo{author}{Zhao, B.} \emph{et~al.}
\newblock \bibinfo{title}{{A Room-Temperature Spin-Valve with van der Waals Ferromagnet Fe5GeTe2/Graphene Heterostructure}}.
\newblock \emph{\bibinfo{journal}{Adv. Mater.}} \textbf{\bibinfo{volume}{35}}, \bibinfo{pages}{2209113} (\bibinfo{year}{2023}).

\bibitem{Ren2011OptimizingRegime}
\bibinfo{author}{Ren, Z.}, \bibinfo{author}{Taskin, A.~A.}, \bibinfo{author}{Sasaki, S.}, \bibinfo{author}{Segawa, K.} \& \bibinfo{author}{Ando, Y.}
\newblock \bibinfo{title}{{Optimizing Bi2-xSbxTe3-ySey solid solutions to approach the intrinsic topological insulator regime}}.
\newblock \emph{\bibinfo{journal}{Phys. Rev. B: Condens. Matter.}} \textbf{\bibinfo{volume}{84}}, \bibinfo{pages}{165311} (\bibinfo{year}{2011}).

\bibitem{Pan2014LowStudy}
\bibinfo{author}{Pan, Y.} \emph{et~al.}
\newblock \bibinfo{title}{{Low carrier concentration crystals of the topological insulator Bi2-xSbxTe3-ySey: A magnetotransport study}}.
\newblock \emph{\bibinfo{journal}{New J. Phys.}} \textbf{\bibinfo{volume}{16}}, \bibinfo{pages}{123035} (\bibinfo{year}{2014}).

\bibitem{Bisswanger2022CVDTemperature}
\bibinfo{author}{Bisswanger, T.} \emph{et~al.}
\newblock \bibinfo{title}{{CVD Bilayer Graphene Spin Valves with 26 {$\mu$}m Spin Diffusion Length at Room Temperature}}.
\newblock \emph{\bibinfo{journal}{Nano Lett.}} \textbf{\bibinfo{volume}{22}}, \bibinfo{pages}{4949–4955} (\bibinfo{year}{2022}).

\bibitem{Fert2001ConditionsSemiconductor}
\bibinfo{author}{Fert, A.} \& \bibinfo{author}{Jaffr{\`{e}}s, H.}
\newblock \bibinfo{title}{{Conditions for efficient spin injection from a ferromagnetic metal into a semiconductor}}.
\newblock \emph{\bibinfo{journal}{Phys. Rev. B: Condens. Matter.}} \textbf{\bibinfo{volume}{64}}, \bibinfo{pages}{184420} (\bibinfo{year}{2001}).

\bibitem{Kamalakar2016InversionHeterostructures}
\bibinfo{author}{Kamalakar, M.~V.}, \bibinfo{author}{Dankert, A.}, \bibinfo{author}{Kelly, P.~J.} \& \bibinfo{author}{Dash, S.~P.}
\newblock \bibinfo{title}{{Inversion of Spin Signal and Spin Filtering in Ferromagnet|Hexagonal Boron Nitride-Graphene van der Waals Heterostructures}}.
\newblock \emph{\bibinfo{journal}{Sci. Rep.}} \textbf{\bibinfo{volume}{6}}, \bibinfo{pages}{21168} (\bibinfo{year}{2016}).

\bibitem{Sinova2015SpinEffects}
\bibinfo{author}{Sinova, J.}, \bibinfo{author}{Valenzuela, S.~O.}, \bibinfo{author}{Wunderlich, J.}, \bibinfo{author}{Back, C.~H.} \& \bibinfo{author}{Jungwirth, T.}
\newblock \bibinfo{title}{{Spin Hall effects}}.
\newblock \emph{\bibinfo{journal}{Rev. Mod. Phys.}} \textbf{\bibinfo{volume}{87}}, \bibinfo{pages}{1213--1260} (\bibinfo{year}{2015}).

\bibitem{Sun2019LargeHeterostructures}
\bibinfo{author}{Sun, R.} \emph{et~al.}
\newblock \bibinfo{title}{{Large Tunable Spin-to-Charge Conversion Induced by Hybrid Rashba and Dirac Surface States in Topological Insulator Heterostructures}}.
\newblock \emph{\bibinfo{journal}{Nano Lett.}} \textbf{\bibinfo{volume}{19}}, \bibinfo{pages}{4420--4426} (\bibinfo{year}{2019}).

\bibitem{Zhao2020UnconventionalWTe2}
\bibinfo{author}{Zhao, B.} \emph{et~al.}
\newblock \bibinfo{title}{{Unconventional Charge–Spin Conversion in Weyl-Semimetal WTe2}}.
\newblock \emph{\bibinfo{journal}{Adv. Mater.}} \textbf{\bibinfo{volume}{32}}, \bibinfo{pages}{2000818} (\bibinfo{year}{2020}).

\bibitem{Zhao2020ObservationTemperature}
\bibinfo{author}{Zhao, B.} \emph{et~al.}
\newblock \bibinfo{title}{{Observation of charge to spin conversion in Weyl semimetal WTe2 at room temperature}}.
\newblock \emph{\bibinfo{journal}{Phys. Rev. Res.}} \textbf{\bibinfo{volume}{2}}, \bibinfo{pages}{013286} (\bibinfo{year}{2020}).

\bibitem{Zhao2020Charge-spinDesign}
\bibinfo{author}{Zhao, B.}, \bibinfo{author}{Hoque, A.~M.}, \bibinfo{author}{Khokhriakov, D.}, \bibinfo{author}{Karpiak, B.} \& \bibinfo{author}{Dash, S.~P.}
\newblock \bibinfo{title}{{Charge-spin conversion signal in WTe2 van der Waals hybrid devices with a geometrical design}}.
\newblock \emph{\bibinfo{journal}{Appl. Phys. Lett.}} \textbf{\bibinfo{volume}{117}}, \bibinfo{pages}{242401} (\bibinfo{year}{2020}).

\bibitem{Hoque2020Charge-spinHeterostructures}
\bibinfo{author}{Hoque, A.~M.}, \bibinfo{author}{Khokhriakov, D.}, \bibinfo{author}{Karpiak, B.} \& \bibinfo{author}{Dash, S.~P.}
\newblock \bibinfo{title}{{Charge-spin conversion in layered semimetal TaTe2 and spin injection in van der Waals heterostructures}}.
\newblock \emph{\bibinfo{journal}{Phys. Rev. Res.}} \textbf{\bibinfo{volume}{2}}, \bibinfo{pages}{033204} (\bibinfo{year}{2020}).

\bibitem{Kovacs-Krausz2020ElectricallyBiTeBr}
\bibinfo{author}{Kov{\'{a}}cs-Krausz, Z.} \emph{et~al.}
\newblock \bibinfo{title}{{Electrically Controlled Spin Injection from Giant Rashba Spin-Orbit Conductor BiTeBr}}.
\newblock \emph{\bibinfo{journal}{Nano Lett.}} \textbf{\bibinfo{volume}{20}}, \bibinfo{pages}{4782--4791} (\bibinfo{year}{2020}).

\bibitem{Ontoso2023UnconventionalHeterostructures}
\bibinfo{author}{Ontoso, N.} \emph{et~al.}
\newblock \bibinfo{title}{{Unconventional Charge-to-Spin Conversion in Graphene/ MoTe2 van der Waals Heterostructures}}.
\newblock \emph{\bibinfo{journal}{Phys. Rev. Appl.}} \textbf{\bibinfo{volume}{19}}, \bibinfo{pages}{014053} (\bibinfo{year}{2023}).

\bibitem{Hoque2022ChargeNbSe2b}
\bibinfo{author}{Hoque, A.~M.}, \bibinfo{author}{Zhao, B.}, \bibinfo{author}{Khokhriakov, D.}, \bibinfo{author}{Muduli, P.} \& \bibinfo{author}{Dash, S.~P.}
\newblock \bibinfo{title}{{Charge to spin conversion in van der Waals metal NbSe2}}.
\newblock \emph{\bibinfo{journal}{Appl. Phys. Lett.}} \textbf{\bibinfo{volume}{121}}, \bibinfo{pages}{242404} (\bibinfo{year}{2022}).

\bibitem{Benitez2020TunableHeterostructures}
\bibinfo{author}{Ben{\'{i}}tez, L.~A.} \emph{et~al.}
\newblock \bibinfo{title}{{Tunable room-temperature spin galvanic and spin Hall effects in van der Waals heterostructures}}.
\newblock \emph{\bibinfo{journal}{Nat. Mater.}} \textbf{\bibinfo{volume}{19}}, \bibinfo{pages}{170--175} (\bibinfo{year}{2020}).

\bibitem{Kamalakar2014EnhancedNitride}
\bibinfo{author}{Kamalakar, M.~V.}, \bibinfo{author}{Dankert, A.}, \bibinfo{author}{Bergsten, J.}, \bibinfo{author}{Ive, T.} \& \bibinfo{author}{Dash, S.~P.}
\newblock \bibinfo{title}{{Enhanced tunnel spin injection into graphene using chemical vapor deposited hexagonal boron nitride}}.
\newblock \emph{\bibinfo{journal}{Sci. Rep.}} \textbf{\bibinfo{volume}{4}}, \bibinfo{pages}{6146} (\bibinfo{year}{2014}).

\bibitem{Khokhriakov2020Gate-tunableTemperature}
\bibinfo{author}{Khokhriakov, D.}, \bibinfo{author}{Hoque, A.~M.}, \bibinfo{author}{Karpiak, B.} \& \bibinfo{author}{Dash, S.~P.}
\newblock \bibinfo{title}{{Gate-tunable spin-galvanic effect in graphene-topological insulator van der Waals heterostructures at room temperature}}.
\newblock \emph{\bibinfo{journal}{Nat. Commun.}} \textbf{\bibinfo{volume}{11}}, \bibinfo{pages}{3657} (\bibinfo{year}{2020}).

\bibitem{Lee2020ExtractionModulation}
\bibinfo{author}{Lee, C.~J.} \emph{et~al.}
\newblock \bibinfo{title}{{Extraction of intrinsic field-effect mobility of graphene considering effects of gate-bias-induced contact modulation}}.
\newblock \emph{\bibinfo{journal}{J. Appl. Phys.}} \textbf{\bibinfo{volume}{127}}, \bibinfo{pages}{185105} (\bibinfo{year}{2020}).

\bibitem{Hoque2021All-electricalTemperature}
\bibinfo{author}{Hoque, A.~M.} \emph{et~al.}
\newblock \bibinfo{title}{{All-electrical creation and control of spin-galvanic signal in graphene and molybdenum ditelluride heterostructures at room temperature}}.
\newblock \emph{\bibinfo{journal}{Commun. Phys.}} \textbf{\bibinfo{volume}{4}}, \bibinfo{pages}{124} (\bibinfo{year}{2021}).

\bibitem{Khokhriakov2018TailoringHeterostructures}
\bibinfo{author}{Khokhriakov, D.} \emph{et~al.}
\newblock \bibinfo{title}{{Tailoring emergent spin phenomena in Dirac material heterostructures}}.
\newblock \emph{\bibinfo{journal}{Sci. Adv.}} \textbf{\bibinfo{volume}{4}}, \bibinfo{pages}{9349} (\bibinfo{year}{2018}).

\bibitem{Hoque2023Spin-valleyHeterostructure}
\bibinfo{author}{Hoque, A.~M.} \emph{et~al.}
\newblock \bibinfo{title}{{Spin-valley coupling and spin-relaxation anisotropy in all-CVD Graphene-MoS2 van der Waals heterostructure}}.
\newblock \emph{\bibinfo{journal}{Phys. Rev. Mater.}} \textbf{\bibinfo{volume}{7}}, \bibinfo{pages}{044005} (\bibinfo{year}{2023}).

\bibitem{Karpiak20181DHeterostructures}
\bibinfo{author}{Karpiak, B.} \emph{et~al.}
\newblock \bibinfo{title}{{1D ferromagnetic edge contacts to 2D graphene/h-BN heterostructures}}.
\newblock \emph{\bibinfo{journal}{2D Mater.}} \textbf{\bibinfo{volume}{5}}, \bibinfo{pages}{014001} (\bibinfo{year}{2018}).

\bibitem{Han2010TunnelingGraphene}
\bibinfo{author}{Han, W.} \emph{et~al.}
\newblock \bibinfo{title}{{Tunneling spin injection into single layer graphene}}.
\newblock \emph{\bibinfo{journal}{Phys. Rev. Lett.}} \textbf{\bibinfo{volume}{105}}, \bibinfo{pages}{167202} (\bibinfo{year}{2010}).

\bibitem{Kamalakar2015LongGraphene}
\bibinfo{author}{Kamalakar, M.~V.}, \bibinfo{author}{Groenveld, C.}, \bibinfo{author}{Dankert, A.} \& \bibinfo{author}{Dash, S.~P.}
\newblock \bibinfo{title}{{Long distance spin communication in chemical vapour deposited graphene}}.
\newblock \emph{\bibinfo{journal}{Nat. Commun.}} \textbf{\bibinfo{volume}{6}}, \bibinfo{pages}{6766} (\bibinfo{year}{2015}).

\bibitem{Yan2017LargeHeterostructure}
\bibinfo{author}{Yan, W.} \emph{et~al.}
\newblock \bibinfo{title}{{Large room temperature spin-to-charge conversion signals in a few-layer graphene/Pt lateral heterostructure}}.
\newblock \emph{\bibinfo{journal}{Nat. Commun.}} \textbf{\bibinfo{volume}{8}}, \bibinfo{pages}{661} (\bibinfo{year}{2017}).

\bibitem{Safeer2019LargeTemperature}
\bibinfo{author}{Safeer, C.~K.} \emph{et~al.}
\newblock \bibinfo{title}{{Large Multidirectional Spin-to-Charge Conversion in Low-Symmetry Semimetal MoTe2 at Room Temperature}}.
\newblock \emph{\bibinfo{journal}{Nano Lett.}} \textbf{\bibinfo{volume}{19}}, \bibinfo{pages}{8758--8766} (\bibinfo{year}{2019}).

\bibitem{Sierra2018ThermoelectricGraphene}
\bibinfo{author}{Sierra, J.~F.} \emph{et~al.}
\newblock \bibinfo{title}{{Thermoelectric spin voltage in graphene}}.
\newblock \emph{\bibinfo{journal}{Nat. Nanotechnol.}} \textbf{\bibinfo{volume}{13}}, \bibinfo{pages}{107--111} (\bibinfo{year}{2018}).

\bibitem{Safeer2019Room-TemperatureHeterostructures}
\bibinfo{author}{Safeer, C.~K.} \emph{et~al.}
\newblock \bibinfo{title}{{Room-Temperature Spin Hall Effect in Graphene/MoS2 van der Waals Heterostructures}}.
\newblock \emph{\bibinfo{journal}{Nano Lett.}} \textbf{\bibinfo{volume}{19}}, \bibinfo{pages}{1074–1082} (\bibinfo{year}{2019}).

\bibitem{Ingla-aynes2022OmnidirectionalHeterostructuresb}
\bibinfo{author}{Ingla-ayn{\'{e}}s, J.} \emph{et~al.}
\newblock \bibinfo{title}{{Omnidirectional spin-to-charge conversion in graphene/NbSe2 van der Waals heterostructures}}.
\newblock \emph{\bibinfo{journal}{2D Mater.}} \textbf{\bibinfo{volume}{9}}, \bibinfo{pages}{045001} (\bibinfo{year}{2022}).

\bibitem{Rodriguez-Vega2017GiantHeterostructures}
\bibinfo{author}{Rodriguez-Vega, M.}, \bibinfo{author}{Schwiete, G.}, \bibinfo{author}{Sinova, J.} \& \bibinfo{author}{Rossi, E.}
\newblock \bibinfo{title}{{Giant Edelstein effect in topological-insulator-graphene heterostructures}}.
\newblock \emph{\bibinfo{journal}{Phys. Rev. B: Condens. Matter.}} \textbf{\bibinfo{volume}{96}}, \bibinfo{pages}{235419} (\bibinfo{year}{2017}).

\bibitem{Li2020Gate-TunableTemperature}
\bibinfo{author}{Li, L.} \emph{et~al.}
\newblock \bibinfo{title}{{Gate-Tunable Reversible Rashba-Edelstein Effect in a Few-Layer Graphene/2H-TaS2 Heterostructure at Room Temperature}}.
\newblock \emph{\bibinfo{journal}{ACS Nano}} \textbf{\bibinfo{volume}{14}}, \bibinfo{pages}{5251--5259} (\bibinfo{year}{2020}).

\bibitem{Song2018SpinHeterostructures}
\bibinfo{author}{Song, K.} \emph{et~al.}
\newblock \bibinfo{title}{{Spin Proximity Effects in Graphene/Topological Insulator Heterostructures}}.
\newblock \emph{\bibinfo{journal}{Nano Lett.}} \textbf{\bibinfo{volume}{18}}, \bibinfo{pages}{2033--2039} (\bibinfo{year}{2018}).

\bibitem{Dieny2020OpportunitiesIndustry}
\bibinfo{author}{Dieny, B.} \emph{et~al.}
\newblock \bibinfo{title}{{Opportunities and challenges for spintronics in the microelectronics industry}}.
\newblock \emph{\bibinfo{journal}{Nat. Electron}} \textbf{\bibinfo{volume}{3}}, \bibinfo{pages}{446--459} (\bibinfo{year}{2020}).

\end{thebibliography}

\end{document}